\documentclass[twocolumn, tighten, dvipsnames, floatfix, linenumbers]{aastex63}

\usepackage{ulem}
\usepackage{float}
\usepackage{gensymb}
\usepackage{amsmath} % or simply amstext
\usepackage[]{xcolor}
\usepackage{booktabs}
\usepackage{physics}
\usepackage{longtable}
\usepackage{array}
\usepackage{changepage}
\usepackage{threeparttable}
\usepackage[flushmargin]{footmisc}
\DeclareUnicodeCharacter{2212}{-}
\usepackage[title]{appendix}
\usepackage[utf8]{inputenc}
\usepackage[T1]{fontenc}
\usepackage{textcomp}
\usepackage{gensymb}
\newcommand{\kms}{km\,s$^{-1}$}
\usepackage[version=4]{mhchem}
\usepackage{lineno}

\nolinenumbers

\usepackage{amsmath}
\usepackage{lipsum}
\usepackage{float}

\shorttitle{The Triangulum-Andromeda Overdensity}
\shortauthors{Abuchaim et al.}

\begin{document}

\title{The Chemodynamical Nature of the Triangulum-Andromeda Overdensity} 

\author[0000-0002-6838-2178]{Yuri Abuchaim}
\affiliation{Universidade de S\~ao Paulo, Instituto de Astronomia, Geof\'isica e Ci\^encias Atmosf\'ericas, Departamento de Astronomia, SP 05508-090, S\~ao Paulo, Brazil}

\author[0000-0002-0537-4146]{H\'elio D. Perottoni}
\affiliation{Universidade de S\~ao Paulo, Instituto de Astronomia, Geof\'isica e Ci\^encias Atmosf\'ericas, Departamento de Astronomia, SP 05508-090, S\~ao Paulo, Brazil}
\affil{Institut de Ciències del Cosmos, Universitat de Barcelona (IEEC-UB), Martí i Franquès 1, E-08028 Barcelona, Spain}

\author[0000-0001-7479-5756]{Silvia Rossi}
\affiliation{Universidade de S\~ao Paulo, Instituto de Astronomia, Geof\'isica e Ci\^encias Atmosf\'ericas, Departamento de Astronomia, SP 05508-090, S\~ao Paulo, Brazil}

\author[0000-0002-9269-8287]{Guilherme Limberg}
\affil{Universidade de S\~ao Paulo, Instituto de Astronomia, Geof\'isica e Ci\^encias Atmosf\'ericas, Departamento de Astronomia, SP 05508-090, S\~ao Paulo, Brazil}
\affil{Department of Astronomy \& Astrophysics, University of Chicago, 5640 S. Ellis Ave., Chicago, IL 60637, USA}
\affil{Kavli Institute for Cosmological Physics, University of Chicago, %5640 S. Ellis Avenue, 
Chicago, IL 60637, USA}

\author[0000-0002-5974-3998]{Angeles P\'erez-Villegas}
\affiliation{Instituto de Astronom\'ia, Universidad Nacional Aut\'onoma de M\'exico, Apartado Postal 106, C. P. 22800, Ensenada, B. C., Mexico}

\author[0000-0002-7529-1442]{Rafael M. Santucci}
\affiliation{Universidade Federal de Goi\'as, Instituto de Estudos Socioambientais, Planet\'ario, Goi\^ania, GO 74055-140, Brazil}
\affiliation{Universidade Federal de Goi\'as, Instituto de F\'isica, Goi\^ania, GO 74001-970, Brazil}

\author[0000-0003-4479-1265]{Vinicius M. Placco}
\affiliation{NSF’s NOIRLab, 950 N. Cherry Ave., Tucson, AZ 85719, USA}

\author[0000-0003-0636-7463]{Jo\~ao V. Sales-Silva}
\affil{Observat\'orio Nacional/MCTIC, R. Gen. Jos\'e Cristino, 77,  20921-400, Rio de Janeiro, Brazil}

\author[0000-0003-4524-9363]{Friedrich Anders}
\affil{Institut de Ciències del Cosmos, Universitat de Barcelona (IEEC-UB), Martí i Franquès 1, E-08028 Barcelona, Spain}

\author[0000-0002-5274-4955]{Helio J. Rocha-Pinto}
\affiliation{Universidade Federal do Rio de Janeiro, Observat\'orio do Valongo, Lad. Pedro Ant\^onio 43, 20080-090, Rio de Janeiro, Brazil}

\bigskip\bigskip\smallskip\smallskip\smallskip\smallskip\smallskip\smallskip\smallskip

\begin{abstract}

We present a chemodynamical study of the Triangulum-Andromeda overdensity (TriAnd) employing a sample of 31 candidate stars observed with the GRACES high-resolution ($R$=40,000) spectrograph at the Gemini North (8.1 m) telescope. TriAnd is a stellar substructure found toward the outer disk of the Milky Way, located at $R_{\rm GC}\sim 18$ kpc from the Sun, toward Galactic latitude $b \sim 25 \degree$. Most stars in our sample have dynamical properties compatible with a disk stellar population. In addition, by applying an eccentricity cut, we are able to detect a stellar contamination that seems to be consistent with an accreted population. In chemical abundance space, the majority of our TriAnd candidates are similar to the outer thin-disk population, suggesting that the overdensity has an \textit{in situ} origin. Finally, the found accreted halo interlopers spatially overlapping with TriAnd should explain the historical discussion of the overdensity's nature due to its complex chemical patterns.

\end{abstract}


\newpage

\section{Introduction} 
\label{sec:intro}

The Milky Way (MW) is teeming with substructures such as stellar streams and overdensities whose origins are still under debate \citep{belokurov2013,helmi2020}. Among the various scenarios for the formation of our Galaxy \citep{els1962,searle1978,tinsley1980}, the bottom-up scenario is the most accepted one, taking into account the formation of these substructures. First suggested by \cite{searle1978}, it is supported by theoretical predictions from the $\Lambda$ Cold Dark Matter ($\Lambda$CDM) cosmological paradigm \citep{spergel2007} and numerical simulations \citep{somerville2015}.

In this hierarchical scenario, ancient mergers of dwarf galaxies with a still-young MW likely caused the formation of substructures from the tidal disruption of these satellites \citep{johnston1998,bullock2005,johnston2008,cooper2010,helmi2011,gomez2013}. These events, however, did not only occur in ancient times; there are also ongoing accretions like the Sagittarius dwarf spheroidal \citep{ibata1994, ibata1995}. Additionally, a profusion of other disrupted dwarf galaxies that populate the Galactic halo can be seen in the form of stellar streams \citep{helmi1999,ivezic2000,yanny2000,belokurov2006,belokurov2007,newberg2009,grillmair2011,bernard2016,grillmair2016,malhan2018,mateu2018,shipp2018,ibata2019,ibata2021,li2022}.

Stellar overdensities are among the different substructures permeating the Galactic halo, e.g. Virgo \citep{newberg2002,juric2008}, Pisces \citep{sesar2007,watkins2009}, Eridanus-Phoenix \citep{li2016} and Hercules-Aquila \citep{belokurov2007b}. Ancient mergers seem to be the main culprits for the formation of these substructures \citep{belokurov2019, donlon2019, chang2020, balbinot2021, naidu2021,chandra2022,helio2022, wang2022}, but they can also be found close to the Galactic plane with a potentially different origin.

Historically, an \textit{in situ} origin for the stellar overdensities close to the Galactic plane has been suggested \citep{rocha_pinto2003, momany2004, momany2006, lopez_corredoira2007, gomez2013, dierickx2014, price2015, xu2015, li2017, li2021, deason2018, sheffield2018, youakim2020, carballo_bello2021, ramos2021, laporte2022}. Those structures are proposed to be formed as a result of interactions of dwarf galaxies with the MW throughout its history, which can dynamically heat the disk \citep{yanny2016, figueras2017, schonrich2018}, supported by simulations \citep{laporte2018, laporte2019}. Other authors, by comparing both the chemical and dynamical patterns from those stellar overdensities with satellite galaxies, conclude that their most probable origin is extragalactic \citep{martin2004, bellazzini2006, chou2011, meisner2012, deason2014, morganson2016, guglielmo2018}.

One of the stellar overdensities close to the Galactic plane is the Triangulum-Andromeda overdensity (TriAnd), first identified by \cite{rocha_pinto2004} using a photometric selection of M-giant star candidates from the Two Micron All Sky Survey (2MASS; \citealt{2MASS}) between $-100\degree < l < -150\degree$ and $-40\degree < b < -20\degree$ as a clumpy cloudlike structure. \cite{majewski2004} also observed TriAnd through a deep photometric survey of M31 fields. Both works estimated a distance of ${\sim}$16--25 kpc from the Sun.

Since its discovery, TriAnd has been studied with different techniques to better understand its characteristics and plausible origin. \cite{martin2014}, utilizing MegaCam photometric data \citep{martin2007} from the Pan-Andromeda Archaeological Survey \citep{pandas_survey}, identified the main sequence of TriAnd stars and other structures also in the same region. \cite{xu2015}, with photometric data from the Sloan Digital Sky Survey (SDSS; \citealt{sdss}), argued that TriAnd is, apparently, a concentric structure located 21 kpc from the Galactic center. \cite{helio2018}, also with SDSS data, observed fluctuations in the structure's density being limited at $b=-45\degree$, in agreement with previous kinematical determinations \citep{sheffield2014}.

Among the different techniques employed to better understand TriAnd's features, spectroscopy has been the most prominent to settle the debate about the nature of the overdensity. In the first high-resolution ($R=$32,000 and 35,000) spectroscopic study of TriAnd, \cite{chou2011} analyzed the spectra of six candidate stars selected from \cite{rocha_pinto2004}. Their results indicated that TriAnd would have higher metallicity ($\rm[Fe/H] = -0.64 \pm{0.08}$ dex) than previously estimated ($\rm[Fe/H] \sim -1.2$ dex) by \cite{rocha_pinto2004} using the Ca infrared triplet spectral indices from lower-resolution ($R\sim3300$) spectra.

Recently, with high-resolution ($R$=36,000 and 47,000) spectroscopy, \cite{bergemann2018} derived abundance patterns for O, Na, Mg, Ti, Ba, and Eu considering TriAnd candidate star samples with mean $\rm[Fe/H] = −0.59 \pm 0.12$ dex. Their main result was that these stars have an \textit{in situ} disk origin and suggested that the overdensity was formed by tidal interactions of the disk with passing or merging dwarf galaxies. Shortly after this, using data from the Apache Point Observatory Galactic Evolution Experiment \citep{majewski2017} Fourteenth Data Release (APOGEE DR14; \citealt{apogeedr14}), \cite{hayes2018} observed that the chemical patterns obtained from the TriAnd stars (Mg, (C+N), K, Ca, Mn, and Ni) appear to be consistent with a metal-poor extrapolation of the outer disk's trend to a larger radius.

\cite{sales2019, sales2020} obtained a sample of 13 TriAnd candidate stars and studied a larger set of elements, including $\alpha$ (Mg, Ca, Si, and Ti), iron-peak (Fe, Ni, and Cr), odd-$Z$ (Al and Na), and neutron-capture elements (Ba and Eu). The authors concluded that the overdensity structure is composed by stars with a unique chemical pattern not corresponding to stars present in either the local Galactic disk or dwarf spheroidal galaxies.

The complex chemical profile observed for the TriAnd stars, combined with the low number of candidates observed with high-resolution spectroscopy, and the scarcity of studies with field stars in the outer disk region are the most relevant limitations to establish the formation history of this overdensity. In this work, we intend to overcome these challenges with a chemodynamical investigation of a larger and updated sample of 31 TriAnd candidate stars, 13 of which are reanalyzed from the sample presented by \citet{sales2019, sales2020}, plus 18 new stars observed over the last 3 yr with the same instrumentation as the previous works.

This paper is outlined as follows. Section \ref{sec:data} describes the sample of TriAnd candidate stars studied in this work. Section \ref{sec:met} describes the methodologies and techniques employed to derive orbital and atmospheric parameters, radial velocities, and chemical abundances. The analyses and discussions of the dynamical and chemical properties of our TriAnd candidate stars are described in Section \ref{sec:discussion}. Finally, Section \ref{sec:conclusion} presents a summary with concluding remarks.

\section{Data}
\label{sec:data}

\begin{figure*}[ht!]
    \includegraphics[width=2\columnwidth]{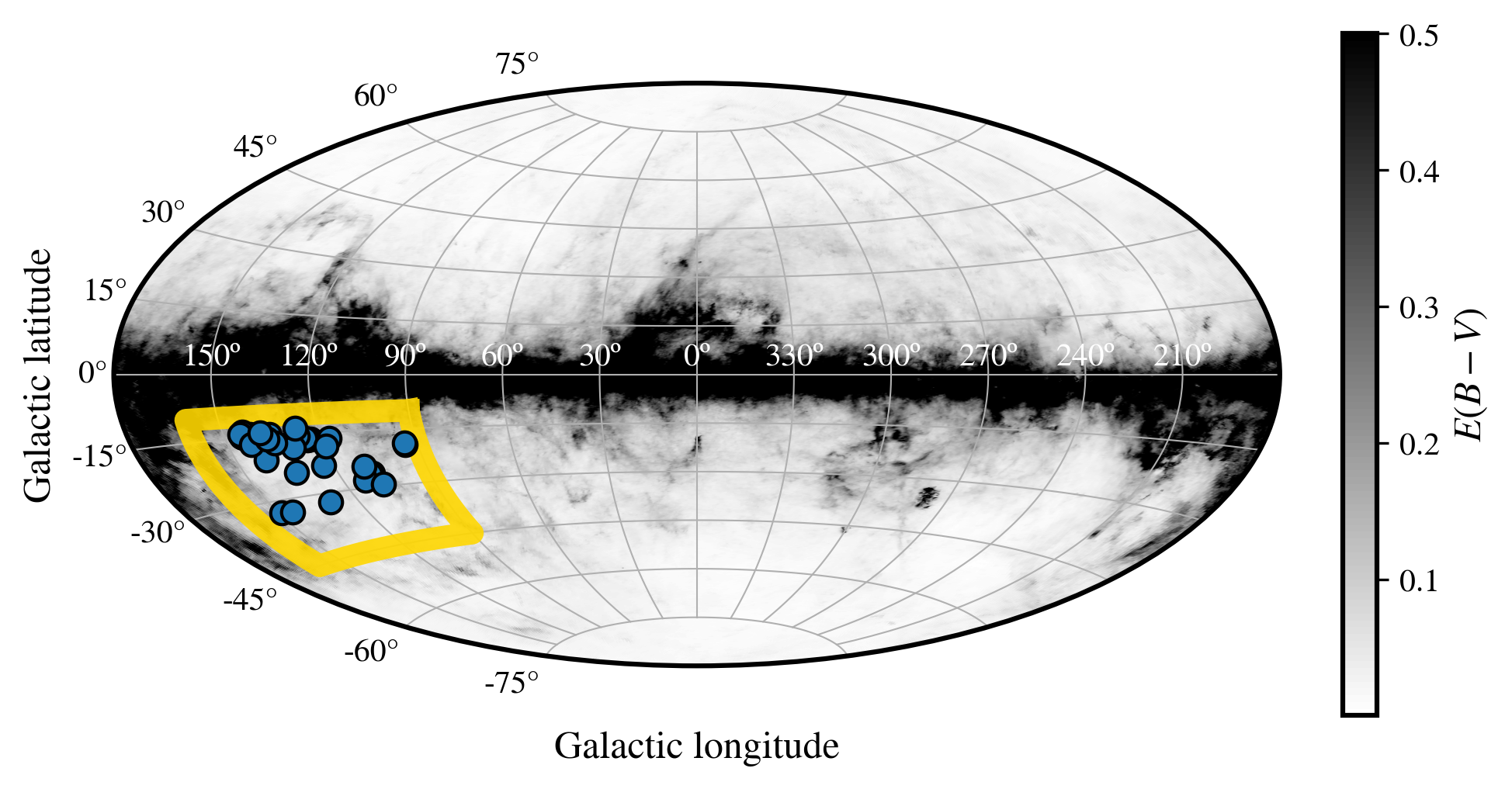}
    \caption{Spatial projection in Galactic coordinates of TriAnd candidate stars employed throughout this work, represented by blue circles. The yellow lines identify the overdensity footprint ($-90\degr < l < -160\degr$ and $-10\degr > b > -45\degr$). The stars are superimposed on the \cite{schlegel1998} interstellar extinction map, while the color bar shows the reddening.} 
    \label{fig:aitoff_xyz}
\end{figure*}

\subsection{Target Selection and Observations}
\label{sec:target}

We selected an M-giant star sample using near-infrared photometric data from the 2MASS catalog \citep{cutri2003}, which was dereddened using the extinction map from \cite{sfd1998} and the extinction laws from \cite{majewski2003}. We adopted the same photometric selection criteria provided by \cite{sales2019}, which efficiently separate M giants from contaminant disk M dwarf stars. This method was successfully used to select distant M giants that led to the discovery, for instance, of Sagittarius tidal tails \citep{majewski2003}, the mapping of the Monoceros overdensity \citep{rocha_pinto2003}, and the discovery of TriAnd \citep{rocha_pinto2004}, as well as spectroscopic follow-up by \cite{sheffield2014}, \cite{bergemann2018}, \cite{hayes2018}, and \cite{sales2019}. The TriAnd candidates were further restricted to the region covering $-90\degr < l < -160\degr$ and $-10\degr > b > -45\degr$, which is associated with TriAnd (\citealt{rocha_pinto2004,Perottoni_2019,laporte2022}).

As a second criterion, we used data from Gaia DR2 \citep{gaiadr2} to select reliable candidates of TriAnd in proper-motion space. As identified by \cite{sales2019}, some sparse stars found in the TriAnd region with $\mu_\alpha^* = \mu_\alpha \cos \delta > 0.5 \ {\rm mas} \ {\rm yr}^{-1}$ have characteristics similar to those from the halo population. In order to avoid contamination by halo interlopers in the region associated with the overdensity, we selected our targets inside an ellipsoid delimited by a $2\sigma$ range around the centroid $(\mu_\alpha^*, \mu_\delta)$ = $(-0.11, -0.57)$ mas yr$^{-1}$ that was estimated with the TriAnd samples of \cite{chou2011}, \cite{sheffield2014} (which contains stars from \citealt{bergemann2018} and \citealt{hayes2018} samples), and the stars classified as TriAnd members from \cite{sales2019}. In Figure \ref{fig:aitoff_xyz}, we present the sky projection in Galactic coordinates of TriAnd candidate stars within the region associated with the overdensity \citep{rocha_pinto2004,Perottoni_2019,laporte2022}.

Our sample consists of 13 TriAnd candidates presented in \cite{sales2019} and 18 new ones, totaling 31 stars. To ensure homogeneity in the observations, we used the same instruments as in \cite{sales2019}. Namely, we observed the 18 TriAnd candidates with the Gemini Remote Access to CFHT ESPaDOnS Spectrograph (GRACES; \citealt{chene2014}), which is connected to the 8.1 m Gemini North telescope on Maunakea in Hawai’i (USA).

All targets were observed in queue mode (GN-2019B-Q-211, GN-2020B-Q-112, GN-2020B-Q-211; PI: Perottoni) in two-fiber (object$+$sky) configuration with resolution ($R\sim$40,000) in the optical region ($4000 < \lambda/$Å $< 10000$). The typical signal-to-noise ratio is $S/N \sim 50{\rm -}70$ pixel$^{-1}$ at 6000 Å. The $S/N$ individual values are listed in Table \ref{table:observ} with the respective target information and exposure time.

\begin{table*}[ht!]
\centering
\caption{TriAnd star candidates employed in This work observed with Gemini/GRACES.}
\begin{tabular}{cccccccccc}
\hline
2MASS ID         & R.A. (deg) & Decl. (deg) & $l$ (deg) & $b$ (deg) & $J$ (mag)   & $H$ (mag)  & $K$ (mag)  & $S/N$ & Exp. (s) \\ \hline
23495808+3445569 & 357.4920 & 34.7658   & 108.8627  & -26.4210  & 11.74 & 10.90 & 10.73 & 56.02 & 880       \\
00534976+4626089 & 13.4573  & 46.4358   & 123.3615  & -16.4339  & 11.91 & 11.03 & 10.87 & 57.52 & 1250      \\
00594094+4614332 & 14.9206  & 46.2426   & 124.4193  & -16.6054  & 12.20 & 11.38 & 11.18 & 61.95 & 1270      \\
...              & ...      & ...       & ...       & ...       & ...   & ...   & ...   & ...   & …         \\ \hline
\end{tabular}
\begin{tablenotes}
        \footnotesize
        \item Only a portion of this table is shown here to demonstrate its form and content. A machine-readable version of the full table is available.
    \end{tablenotes}
\label{table:observ}
\end{table*}

\section{Methodology}
\label{sec:met}
\subsection{Orbital Parameters}

The spectrophotoastrometric heliocentric distances for our sample were estimated with the Bayesian isochrone-fitting code \texttt{StarHorse} \citep{starhorse2, queiroz2020} based on our derived atmospheric parameters. \texttt{StarHorse} did not present a solution for the distance estimation of two stars from our sample. These stars were removed from our analysis in Section \ref{subsec:dyn}.

We calculated the orbits of our star sample with the publicly available Python library \texttt{AGAMA} \citep{agama} for 5 Gyr forward. The Galactic potential model employed is described in \citet{McMillan2017}. We adopted values for the solar Galactocentric distance $R_{\odot}=8.2$ \citep{hawthorn2016}, the local circular velocity of $v_c = 232.8$ \kms \citep{McMillan2017}, and the solar motion with respect to the local standard of rest $(U_{\odot}, V_{\odot}, W_{\odot}) = (11.10,12.24,7.25)$ \kms \citep{schonrich2010}.

For each star, we performed 1000 Monte Carlo realizations of the orbit according to Gaussian distributions of its uncertainties in distance, proper motion, and radial velocity. The medians of the resulting distributions of the dynamical parameters of interest were taken as our nominal values with 16th and 84th percentiles as associated uncertainties.

\subsection{Spectroscopic Analysis}

\subsubsection{Data reduction}

All of the data were reduced with the \texttt{OPERA} pipeline \citep{martioli2012}, which includes bias subtraction, flat-field correction, and wavelength calibration. The IRAF \citep{iraf} package was employed for the spectral normalization. Radial velocities were obtained by cross-correlation against synthetic spectra from \cite{munari2005} using IRAF's \texttt{fxcor} task.

\subsubsection{Atmospheric parameters}
\label{subsec:atmos}

To calculate the atmospheric parameters and chemical abundances for our TriAnd candidate stars, we employed the spectrum synthesis code \texttt{MOOG}\footnote{\url{http://www.as.utexas.edu/~chris/moog.html}} \citep{moog, sobeck2011} and the $\mathtt{q^2}$ (\texttt{qoyllur-quipu}\footnote{\url{https://github.com/astroChasqui/q2}}) Python package \citep{q2}. The radial velocities and the results of the estimated atmospheric parameters can be found in Table \ref{table:atm}.

\begin{table*}[ht!]
\centering
\caption{Derived Atmospheric Parameters and Radial Velocities for TriAnd star candidates employed in This work.}
\begin{tabular}{ccccccccc}
\hline
2MASS ID         & $T_{\rm eff}$ (K) & $\sigma_{T_{\rm eff}}$ (K) & $\log g$ & $\sigma_{\log g}$ & vt (\kms) & $\sigma_{vt}$ (\kms) & RV (\kms) & $\sigma_{\rm RV}$ (\kms)\\ \hline
23495808+3445569 & 3931 & 46              & 1.30   & 0.22            & 1.94 & 0.10          & -21.76  & 0.73          \\ 
00534976+4626089 & 3891 & 73              & 1.08   & 0.19            & 1.77 & 0.11          & -211.68 & 0.62          \\ 
00594094+4614332 & 4044 & 29              & 0.34   & 0.17            & 1.80 & 0.08          & -131.08 & 0.75          \\ 
... & ... & ... & ... & ... & ... & ... & ... & ... \\ \hline
\end{tabular}
    \begin{tablenotes}
        \footnotesize
        \item Only a portion of this table is shown here to demonstrate its form and content. A machine-readable version of the full table is available.
    \end{tablenotes}
\label{table:atm}
\end{table*}

The atmospheric parameters, $v_t$, $\log g$, and $T_{\rm eff}$, where $v_t$ is the microturbulent velocity, $\log g$ is the logarithm of the surface gravity, and $T_{\rm eff}$ is the effective temperature, along with the metallicity, were obtained from the equivalent widths of the Fe lines. The $T_{\rm eff}$ is calculated from the excitation equilibrium and $\log g$ from the ionization equilibrium between the \ion{Fe}{1} and \ion{Fe}{2} lines. The microturbulent velocity is obtained from the independence between the \ion{Fe}{1} abundances and reduced equivalent width. Finally, [Fe/H] is derived from the ionization equilibrium obtained under the local thermodynamic equilibrium (LTE) atmosphere model.

\begin{figure}[ht!]
    \includegraphics[width=1\columnwidth]{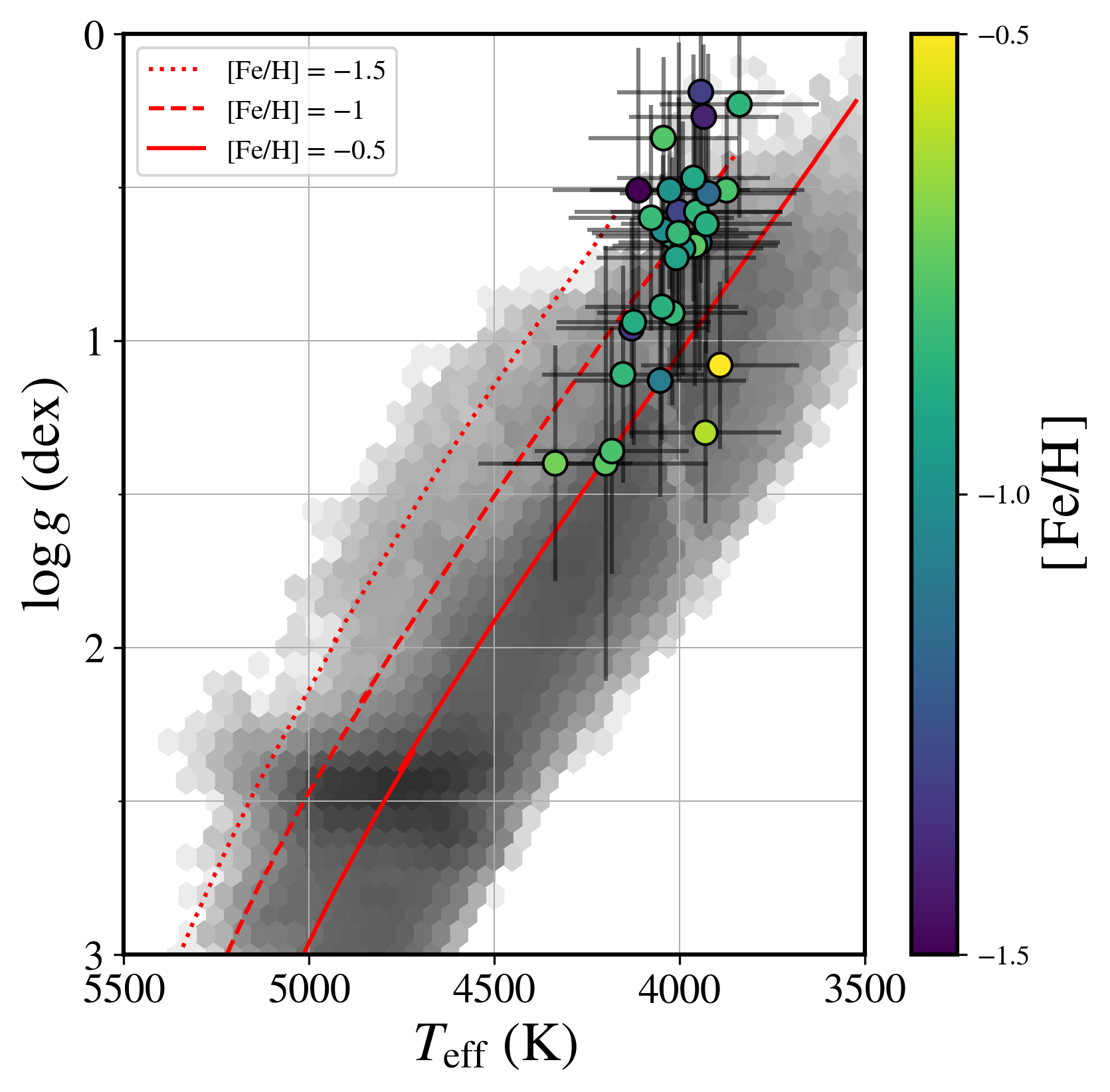}
    \caption{Stellar parameters $\log g$ and $T_{\rm eff}$ obtained for the TriAnd candidate stars analyzed in this work color-coded by metallicity. The error bars indicate the corresponding uncertainties. Field stars from the APOGEE DR17 database are shown as the gray-scale background. Isochrones with 8 Gyr, $[\alpha/Fe]=0$ and varying metallicities from the Darthmouth Stellar Evolution Database are overlaid in red.}
    \label{fig:logteff}
\end{figure}

In Figure \ref{fig:logteff}, we show the respective $\log g$ and $T_{\rm eff}$ values for each star in our sample with overlaid isochrones from the Darthmouth Stellar Evolution Database \citep{darthmouth}. Our obtained atmospheric parameters show that the criteria presented in Section \ref{sec:data} correctly selected M-giant stars. It is worth noting that one common obstacle when working with low-$T_{\rm eff}$ stars in spectroscopy, like M giants, is the presence of molecular bands that can affect the chemical abundance calculations. However, the M-giant stars in our sample are on the upper $T_{\rm eff}$ limit for M-type stars, almost on K-type classification, and the presence of molecular bands is not strong enough to affect the precision of our abundance analyses.

\subsubsection{Chemical abundances}
\label{subsec:elemen}

We determined the chemical abundances of 11 elements for our stars: Na, Mg, Al, Si, Ca, Ti, Cr, Fe, Ni, Ba, and Eu. We employed the line list presented in \cite{sales2019}, with additional lines from the Southern Stellar Stream Spectroscopic Survey \citep{S5} and GALactic Archaeology with HERMES (GALAH; \citealt{GALAH_DR3}) survey, due to the low number of nonblended absorption lines in Mg, Si, Ca, and Na. The line list with the respective measured equivalent widths can be found in Table \ref{table:eqw}.

\begin{table*}[ht!]
\centering
\caption{Equivalent widths for TriAnd star candidates employed in This work.}
\begin{tabular}{cccccccc}
\cline{5-8}
           &         &     &       & \multicolumn{4}{c}{2MASS ID}                                 \\ \hline
Wavelength (Å) & Element & EP (eV)  & $\log gf$   & 23495808+3445569 & 00534976+4626089 & 00594094+4614332 & ... \\ \hline
5682.63    & \ion{Na}{1}  & 2.1 & $-0.7$  & 127.2            & 149.1            & -                & ... \\
5688.2     & \ion{Na}{1}  & 2.1 & $-0.4$  & 133.5            & 149              & -                & ... \\
6154.22    & \ion{Na}{1}  & 2.1 & $-1.51$ & 54.6             & 91.4             & 71.1             & ... \\
...        & ...     & ... & ...   & ...              & ...              & ...              & ... \\ \hline
\end{tabular}
    \begin{tablenotes}
        \footnotesize
        \item EP - excitation potential; $\log gf$ - transition probability.
        \item Only a portion of this table is shown here to demonstrate its form and content. A machine-readable version of the full table is available.
    \end{tablenotes}
\label{table:eqw}
\end{table*}

The equivalent widths of the absorption lines in the analyzed spectra were measured with IRAF's \texttt{splot} task corresponding to each element measured star-by-star on a line-by-line basis. These equivalent widths were then utilized to derive the chemical abundances with \texttt{MOOG} wrapped by the $\mathtt{q^2}$ code. As the resulting abundances are given in absolute values, we employed \cite{grevesse2007} solar values to calculate the corresponding abundances in reference to the Sun. All of the calculations were done assuming LTE with ODFNEW \citep{castelli2003} atmosphere models.

Due to the hyperfine structure and contribution of different isotopes for the neutron-capture elements Ba (lines 5853, 6141, and 6496 Å) and Eu (line 6645 Å), we employed the spectral synthesis method for these elements instead, also with \texttt{MOOG} \citep{sneden2008}. The results of the derived abundances can be found in Table \ref{table:chem}.

\begin{table*}[ht!]
\centering
\caption{Derived Abundances for TriAnd star candidates employed in This work.}
\begin{tabular}{cccccccccc}
\hline
2MASS ID  & {[}Fe/H{]}$_{LTE}$ & {[}Fe/H{]}$_{NLTE}$ & $\sigma_{{[}Fe/H{]}}$ & {[}Na/H{]}$_{LTE}$ & $\sigma_{{[}Na/H{]}}$ & {[}Mg/H{]$_{LTE}$} & {[}Mg/H{]}$_{NLTE}$ & $\sigma_{{[}Mg/H{]}}$   & ... \\ \hline
23495808+3445569 & $-0.56$ &  $-0.55$    & 0.08         & $-0.93$      & 0.08     & $-0.57$ &  $-0.57$  & 0.06    & ... \\
00534976+4626089 & $-0.39$ & $-0.38$     & 0.10      & $-0.53$      & 0.10     & $-0.47$   & $-0.47$  & 0.08      & ... \\
00594094+4614332 & $-0.72$ & $-0.70$    & 0.05    & $-0.55$       & 0.04     & $-0.59$    & $-0.56$  & 0.09   & ... \\
...      & ...        & ... &  ...    & ...        & ...        & ...   & ...    & ...    & ... \\ \hline
\end{tabular}
    \begin{tablenotes}
        \footnotesize
        \item Only a portion of this table is shown here to demonstrate its form and content. A machine-readable version of the full table is available.
    \end{tablenotes}
\label{table:chem}
\end{table*}

As discussed in \cite{q2}, the errors related to the stellar parameters calculated by the $\mathtt{q^2}$ code consider the relationship between the parameters fulfilling the equilibrium conditions as described by \cite{epstein2010} and \cite{bensby2014}. For $[Fe/H]$, the formal error is computed by propagating the errors from the other atmospheric parameters by adding them in quadrature and including the standard error of the mean line-to-line [Fe/H] abundance. For the other chemical abundances, the errors are calculated taking into account both the estimated observed errors, given by the standard deviation from the mean abundance calculated for each set of measured lines, and the error introduced by each atmospheric parameter. The radial velocity errors employing IRAF's \texttt{fxcor} task are computed based on the fitted peak height and the antisymmetric noise, as described in \cite{tonry1979}.

The radial velocities and metallicity ($[Fe/H]$) calculated for our sample have a typical uncertainty of $\pm$0.91 \kms and $\pm$0.11 dex, respectively. The total error budget for the atmospheric parameters was derived by adding in quadrature the statistical errors calculated by the $\mathtt{q^2}$ code with adopted systematic errors of $\Delta T_{\rm eff} = 200$ K and $\Delta \log g = 0.2$. The average uncertainty for each atmospheric parameter is as follows: $T_{\rm eff}$, $\pm$216 K; $\log g$, $\pm$0.39 dex; and $vt$, $\pm$0.19 \kms.

The derived chemical abundances for our sample with the equivalent-width method have a typical uncertainty of $\pm$0.14 dex. The average uncertainty for each element is as follows: \ion{Na}{1}, $\pm$0.16 dex; \ion{Mg}{1}, $\pm$0.11 dex; \ion{Al}{1}, $\pm$0.10 dex; \ion{Si}{1}, $\pm$0.15 dex; \ion{Ca}{1}, $\pm$0.17 dex; \ion{Ti}{1}, $\pm$0.18 dex; \ion{Cr}{1}, $\pm$0.17 dex; \ion{Ni}{1}, $\pm$0.11 dex.

To account for possible departures from LTE for our derived abundances, we performed NLTE corrections for our sample using spectral models from \url{https://nlte.mpia.de}. Based on the studies from \cite{mashonkina2007}, \cite{bergemann2010}, and \cite{bergemann2012,bergemann2013,bergemann2017}, we applied NLTE corrections for Fe, Mg, Ca, Cr, and Si, correcting our derived LTE abundances line by line. We explored NLTE corrections for Ti, but, as observed by \cite{bergemann2018}, NLTE models for Ti do not give consistent solutions with one-dimensional hydrostatic models \citep{bergemann2011}. The average NLTE departure for each element is as follows: \ion{Mg}{1}, $-0.01$ dex; \ion{Si}{1}, $-0.02$ dex; \ion{Ca}{1}, $0.03$ dex; \ion{Cr}{1}, $0.06$ dex; \ion{Fe}{1}, $0.02$ dex. The corresponding NLTE-corrected value of each element described can be found in Table \ref{table:chem}.

\section{Discussions}
\label{sec:discussion}

\subsection{Orbital Parameters}
\label{subsec:dyn}

The history of spectroscopic studies of TriAnd has been controversial since its discovery, with completely opposite results appearing in the literature; \cite{chou2011} originally argued for an extragalactic origin for the overdensity,  but \cite{bergemann2018} and \cite{hayes2018} suggested an \textit{in situ} nature. With the help of the Gaia mission \citep{GaiaMission}, we can now complement the spectroscopic results with the kinematic/dynamical counterpart, allowing us to better understand the complexity of this overdensity.

\begin{figure*}[ht!]
    \centering
    \includegraphics[width=2\columnwidth]{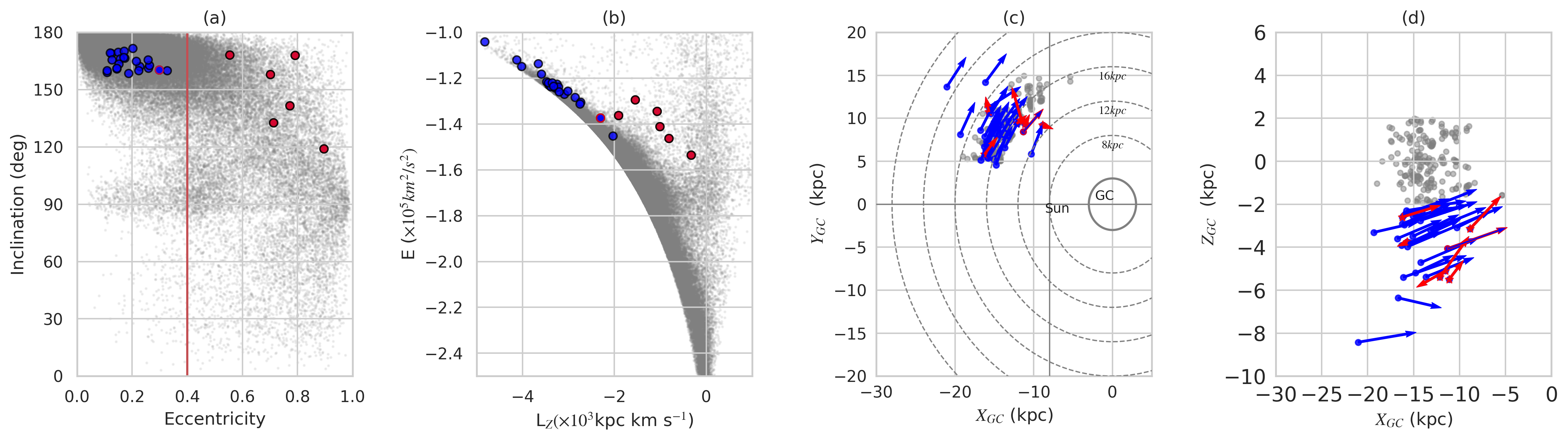}
    \caption{(a) Orbital eccentricity vs. inclination; (b) $(E, L_Z)$; (c) and (d) $(X_{\rm GC},Y_{\rm GC})$ and $(X_{\rm GC},Z_{\rm GC})$ projections, respectively, with vectors representing their velocities in the $X$, $Y$, and $Z$ Galactocentric Cartesian directions. The orbital parameters of TriAnd candidate stars employed in this work are represented in blue. The red circles represent the stars classified with a possible accreted origin. The kinematic criterion applied here is represented by a red line in panel (a). The APOGEE DR17 sample is shown in gray for comparison; panels (a) and (b) show the full sample cleaned of globular cluster and dwarf galaxy stars, and panels (c) and (d) show our restricted outer disk selection ($5 < Y_{\rm GC}/{\rm kpc} < 15$, $\abs{Z_{\rm GC}} < 2 \ {\rm kpc}$, and $R_{\rm GC} > 15 \ {\rm kpc}$).}
    \label{fig:EJs}
\end{figure*}

The orbital parameters of the stars studied in this work, as well as their spatial projections and velocity vectors, are shown in Figure \ref{fig:EJs}. In panel (a), we present the orbital eccentricity vs. inclination\footnote{inclination = $\arccos{(L_z / L)}$, where $L$ is the total angular momentum}. We notice that most of our stars exhibit low eccentricity, which is characteristic of the Galactic disk(s) \citep{freeman2002,kruit2011,hawthorn2016}. We classify those stars as representative of TriAnd, which will be represented in blue for the remainder of this paper. We also included, as gray circles, stars from the APOGEE DR17 database \citep{apogeedr17} for comparison but cleaned of globular cluster and dwarf galaxy stars following \cite{limberg2022}. 

A contamination of high-eccentricity stars in our sample was also detected. Stars on highly eccentric orbits are typically associated with either the \textit{in situ} \citep{dimatteo2019,gallart2019,belokurov2020,bonaca2020} or accreted \citep{koppelman2018,mackereth2019,naidu2020,limberg2021a,myeong2022} halo. As will be discussed in Section \ref{subsec:chem}, these probable halo interlopers also have chemical abundance patterns typical of halo populations \citep{nissen1,nissen2,hayes2018b}.

In order to select the above-described high-eccentricity contamination, we applied a cut in eccentricity to our sample ($e > 0.4$), represented in red throughout this work. This selection classified six stars from our sample (red circles) as possible contamination of \textit{ex situ} stars at the same distance and location as TriAnd.

Figure \ref{fig:EJs}(b), total orbital energy ($E$) versus the $z$-component of the angular momentum ($L_z$), often employed to characterize MW dynamical groups (e.g., \citealt{helmi1999, beers2000, gomez2010, helmi2018, myeong2018}), reinforces that part of those stars are possibly accreted because they have smaller $L_z$ compared to the TriAnd/disk ones (blue circles/gray dots) at the same $E$, a typical characteristic of accreted populations (e.g., \citealt{naidu2020}). On the other hand, TriAnd candidates present $E$ and $L_z$ overlapping with the region of the plot occupied by the outer disk, indicating a common \textit{in situ} origin. 

In Figure \ref{fig:EJs}(c) and (d), we show the $X_{\rm GC}$ vs. $Y_{\rm GC}$ and $X_{\rm GC}$ vs. $Z_{\rm GC}$ projections, respectively. We observe that the velocity vectors from the majority of the sample, in blue, show a corotation with the Galactic disk. However, the red vectors, representing the accreted candidates in our sample, appear to be randomly distributed and do not follow the direction of the Galactic disk, as expected for \textit{ex situ} populations. Our analysis indicates that the majority of the TriAnd members present \textit{in situ}-like orbits, in agreement with other works from the literature \citep{bergemann2018, hayes2018}.

\subsection{Metallicity Distribution Function}
\begin{figure}[ht!]
    \centering
    \includegraphics[width=1\columnwidth]{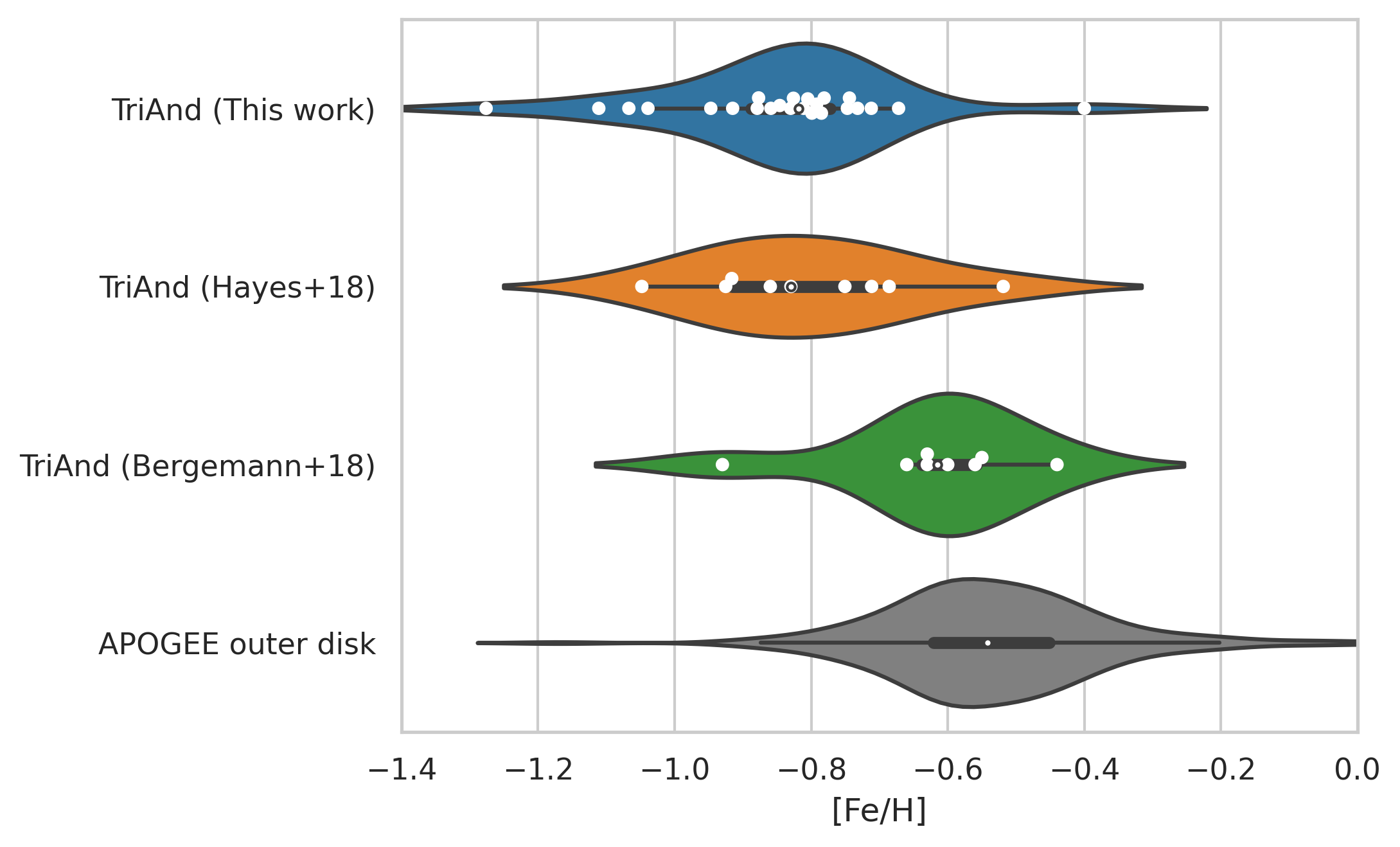}
    \caption{The MDFs for our selected APOGEE outer disk sample and TriAnd stars from the samples of this work, \cite{hayes2018}, and \cite{bergemann2018}. The violin plots represent the distribution shape of each sample, the white dots are the respective metallicity values, the mean value is represented by a black circle with a centered white dot, the thick bar is the interquartile range, and the thin bar represents the 95\% confidence interval.}
    \label{fig:mdf_triand}
\end{figure}

In order to understand the metallicity distribution function (MDF) of the TriAnd population, we analyzed and compared the $[Fe/H]$ derived from this work with other recent spectroscopic analyses. The stars classified as TriAnd in our sample have a  mean metallicity of $[Fe/H] \sim -0.8$ dex, a similar value to that found by the \cite{hayes2018} and \cite{sales2019} samples but more metal-poor than the \cite{bergemann2018} sample. The MDFs for different TriAnd samples are shown in Figure \ref{fig:mdf_triand}.

This mean metallicity value indicates that, on average, our TriAnd candidates are a more metal-poor stellar population than our selected APOGEE outer disk sample. We speculate that the TriAnd population was probably dynamically heated before enriching, maintaining a lower mean metallicity. In this scenario, the younger stellar populations were not dynamically heated enough to reach a $Z_{GC}$ similar to TriAnd, thus explaining the lower mean metallicity in comparison to the outer disk.

As we can see from Figure \ref{fig:mdf_triand}, the \cite{bergemann2018} sample stands out with a higher mean metallicity population when compared to other derived TriAnd samples, with an offset of $\sim 0.2$ dex from our sample. We checked if this feature could be attributed to different sample distances, where stars with a lesser $R_{\rm GC}$ are expected to be more metal-rich. TriAnd candidates from our sample are distributed in the range $12 < R_{\rm GC} < 25$ kpc, whereas most of our stars are found in the range $16 < R_{\rm GC} < 20$ kpc. The \cite{bergemann2018} sample is found at a similar distance ($R_{\rm GC} = 18 \pm 2$ kpc), excluding this alternative. 

We suggest that one of the most plausible explanations for this difference in metallicity found in \cite{bergemann2018} is the method employed by the authors to estimate the atmospheric parameters for their sample, where $T_{\rm eff}$ was estimated by combining the infrared flux method with photometric data from the AAVSO Photometric All-Sky Survey and 2MASS. The use of photometry combined with spectroscopy can lead to an overall higher derived value for ${\rm [Fe/H]}$ when compared to a purely spectroscopic analysis. The use of NLTE models by \cite{bergemann2018} is another possible explanation behind the differences, as both this work and \cite{hayes2018} employ LTE models.

\subsection{Chemical abundances}
\label{subsec:chem}

As we can note from Section \ref{subsec:dyn}, the analysis of the orbital parameters of our TriAnd candidates indicates that they display characteristics of a mixture between \textit{in situ} and accreted stars. With the benefit of chemical analysis, we expect that a more holistic chemodynamical approach can enlighten our understanding of the nature of TriAnd.

Throughout this section, APOGEE DR17 data are utilized for the comparison between our derived TriAnd abundances and MW's accreted and \textit{in situ}, in particular the outer disk, populations. To select outer disk stars, we employed the following criteria: $5 < Y_{\rm GC}/{\rm kpc} < 15$, $\abs{Z_{\rm GC}} < 2 \ {\rm kpc}$, and $R_{\rm GC} = \sqrt{X_{\rm GC}^2 + Y_{\rm GC}^2}$ $> 15 \ {\rm kpc}$. These criteria are more conservative than those applied by \cite{hayes2018} for APOGEE DR14 \citep{apogeedr14} and hence more representative of the outer disk population. To check if TriAnd shares a chemical composition with accreted objects, we use the chemodynamical criteria from \cite{limberg2022}, which were constructed for APOGEE-available abundances and designed to yield minimal contamination from \textit{in situ} stars, to select Gaia-Sausage/Enceladus (GSE; \citealt{belokurov2018, haywood2018}, also \citealt{helmi2018}) members.

\begin{figure*}[ht!]
    \centering
    \includegraphics[width=2\columnwidth]{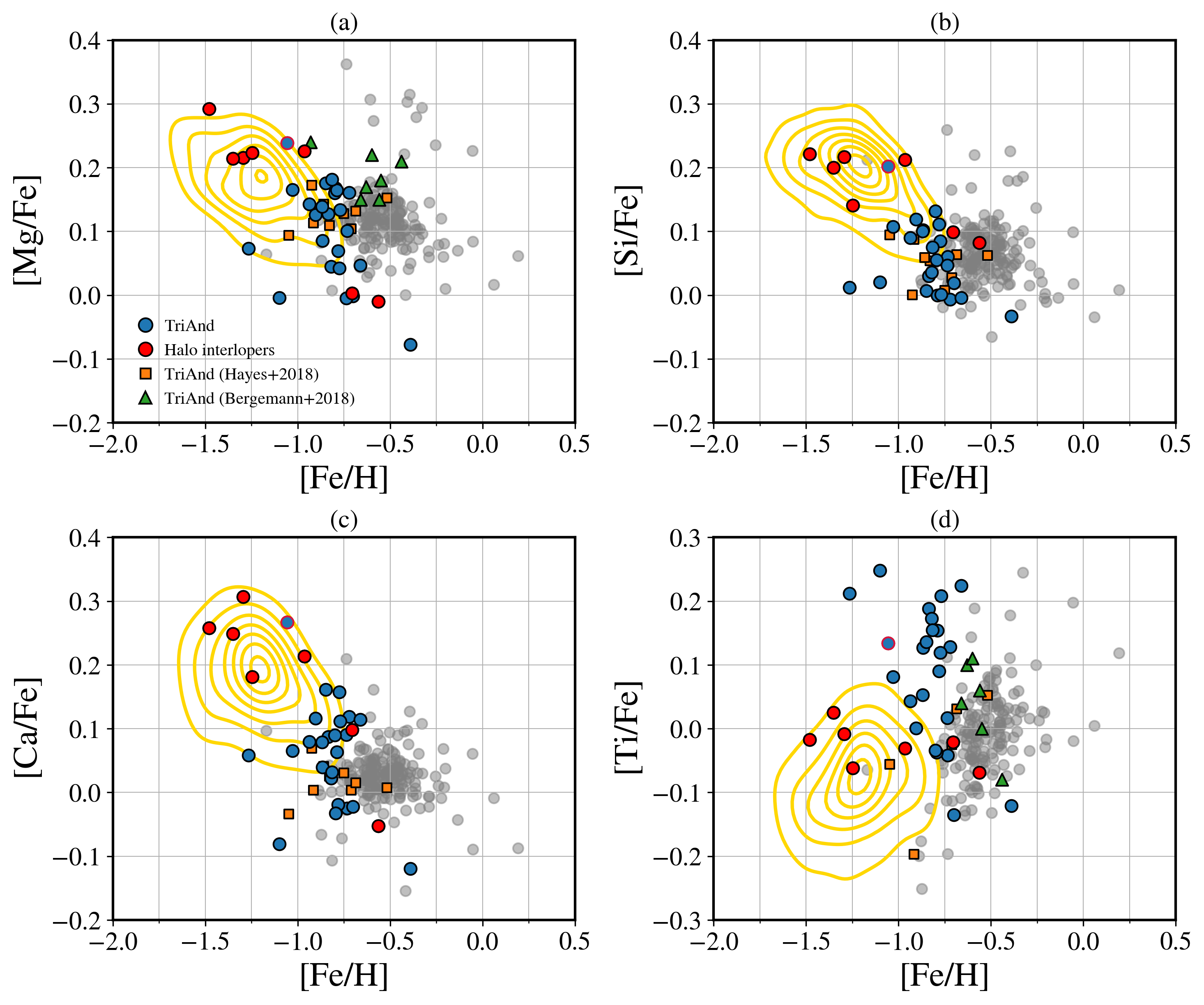}
    \caption{[X/Fe] ratio vs. [Fe/H] for the $\alpha$ elements: (a) Mg,  (b) Si, (c) Ca, and (d) Ti. The TriAnd candidate stars analyzed in this work are represented by blue circles, whereas the possible accreted stars are represented by red ones. The \citet{hayes2018} and \citet{bergemann2018} TriAnd samples are represented by orange squares and green triangles, respectively. Outer disk stars from the APOGEE DR17 database \citep{apogeedr17} are shown in gray. Isodensity contours associated with GSE are represented by yellow lines.}
    \label{fig:alfa_apo_gse}
\end{figure*}

\begin{figure*}[ht!]
    \centering
    \includegraphics[width=2\columnwidth]{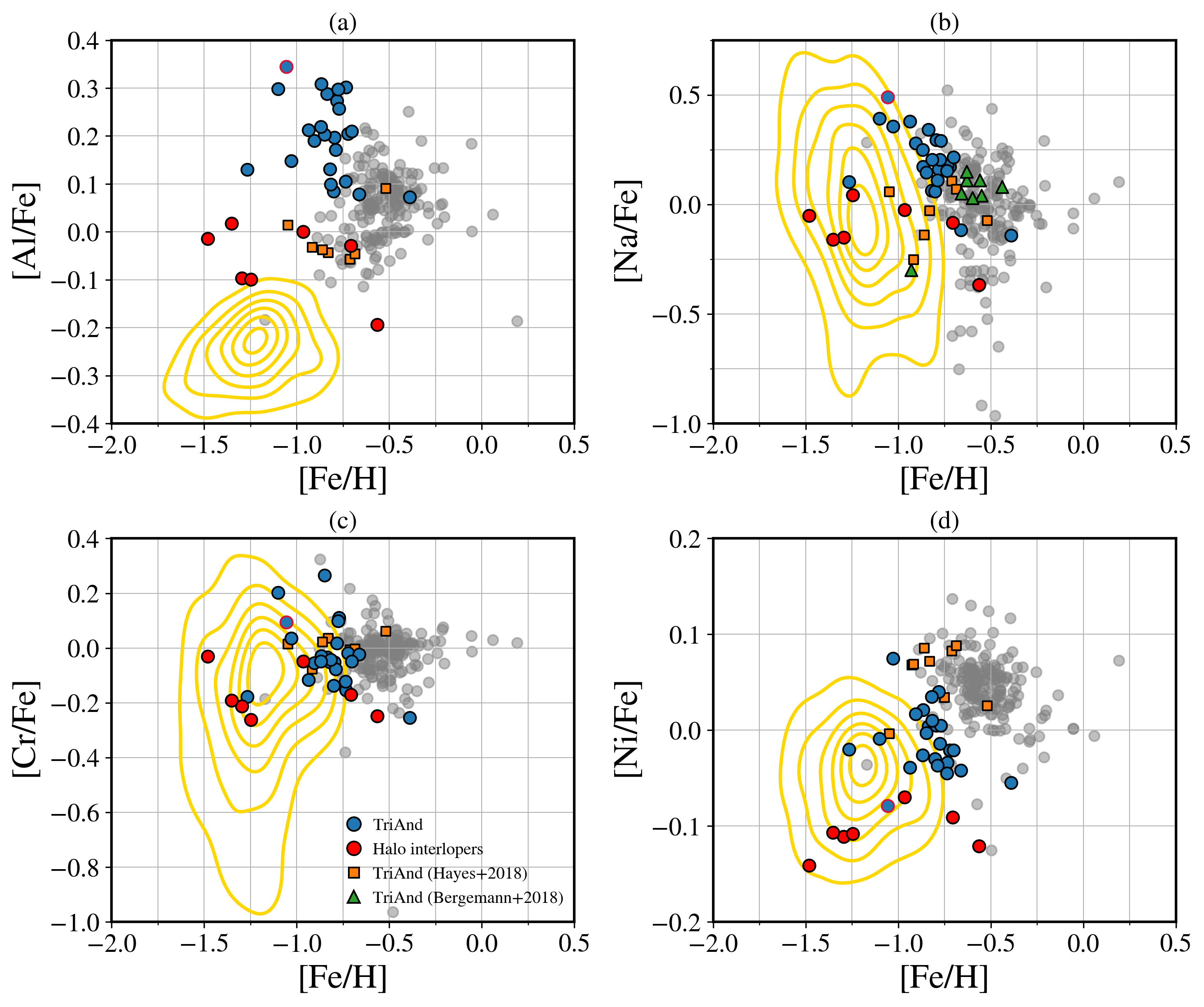}
    \caption{The [X/Fe] ratio vs. [Fe/H] for the odd-$Z$ and iron-peak elements: (a) Al, (b) Na, (c) Cr, and (d) Ni. The TriAnd candidate stars employed in this work are represented by blue circles, whereas the possible accreted stars are represented by red circles. The \cite{hayes2018} and \cite{bergemann2018} samples are represented by orange squares and green triangles, respectively. Outer disk stars from the APOGEE DR17 database \citep{apogeedr17} are shown in gray, and isodensity contours associated with GSE are shown in yellow.}
    \label{fig:nonalfa_apo_gse}
\end{figure*}

In Figures \ref{fig:alfa_apo_gse} and \ref{fig:nonalfa_apo_gse}, we show the estimated abundances for $\alpha$ (Mg, Si, Ca, and Ti), odd-$Z$ (Al and Na), and iron-peak (Cr and Ni) elements for our TriAnd sample. For comparison, we also show the results from other TriAnd spectroscopic studies, namely, \cite{bergemann2018} and \cite{hayes2018}, as well as the aforementioned APOGEE outer disk sample. For consistency with APOGEE DR17, we choose to represent the NLTE-corrected values only for Mg and Ca abundances.

The abundances of Na, Al, Si, Ti, and Cr are not included in \citeauthor{hayes2018}'s (\citeyear{hayes2018}) analysis. Since the authors used APOGEE data, we cross-matched their sample with APOGEE DR17 to obtain the missing chemical abundances and update the Mg, Ca, and Ni values. In addition, \cite{bergemann2018} did not analyze the abundance of Al, Si, Ca, Cr, or Ni. Therefore, they are not included in our comparison. For our interpretations, we also consider the classification into either \textit{in situ} or accreted origin from the eccentricity selection presented in Section \ref{subsec:dyn} and verify if chemical abundances confirm (or reject) our initial conjecture that a mixture of populations can be found at TriAnd's location.

\subsubsection{\texorpdfstring{$\alpha$ elements}{alpha elements}}

The $\alpha$ elements (Mg, Si, Ca, and Ti) are mainly formed in explosive events such as Type II supernovae (SNe II) in high-mass stars \citep{weaver95}. These events occur on a shorter timescale ($\sim$Myr) when compared to other enrichment sources such as Type Ia supernovae (SNe Ia; $\sim$Gyr), formed by merging white dwarf binaries, which are the main source for the iron-peak elements (Cr, Ni, and Fe; \citealt{iwamoto1999}). Therefore, the [$\alpha$/Fe] vs. [Fe/H] relation of a stellar population provides useful information about the relative contribution of SNe II and Ia to the interstellar medium where these stars were formed \citep{matteucci1990}.

In Figure \ref{fig:alfa_apo_gse}, we can see that the stars kinematically classified as part of TriAnd follow a behavior expected for \textit{in situ} stars. The [$\alpha$/Fe] ratio for this TriAnd population is similar to the APOGEE outer disk sample as well as the \cite{bergemann2018} and \cite{hayes2018} TriAnd samples. This behavior can be better appreciated in the chemical abundances of Mg, Si, and Ca, even with an overall low metallicity that sometimes overlaps with the GSE footprint.

The stars classified as likely having an accreted origin fall on top of the contours associated with GSE, showing a higher [$\alpha$/Fe] ratio (${\gtrsim}0.2$ dex) when compared to the thin-disk population. If we look at our full sample, ignoring the eccentricity selection, a ``knee'' pattern can be identified in the relation between the [$\alpha$/Fe] ratio and [Fe/H], as suggested in the literature \citep{sales2020}. 

One star, also studied by \cite{chou2011}, stands out in our sample. Represented by a blue circle outlined in red in Figures \ref{fig:EJs}, \ref{fig:alfa_apo_gse}, \ref{fig:nonalfa_apo_gse}, and \ref{fig:baeu_saga}, this star is not marked as being accreted when taking into account the eccentricity selection method employed, while showing an accreted chemical pattern. As we can observe for this star in particular, these apparent \textit{ex situ} chemical characteristics from the \cite{chou2011} sample could have led the authors to suggest an accreted origin for TriAnd. Given that this star is located at $R_{\rm GC}$ = 14 kpc and based on its chemistry, we speculate that it belongs to the thick/splashed disk population.

\subsubsection{Odd-\texorpdfstring{$Z$}x and Iron-peak Elements}

In Figure \ref{fig:nonalfa_apo_gse}, we show the odd-$Z$ (Al and Na) and iron-peak (Cr and Ni) elements. Odd-$Z$, as for the $\alpha$ elements, are majorly synthesized by the evolution of massive stars but by different nucleosynthesis processes, often present in the red giant branch and asymptotic giant branch (AGB) phases of stellar evolution \citep{herwig2005,ventura2011, depalo2016}. These different processes are reflected in the stellar population of the MW, where a clear difference between the abundances of odd-$Z$ and $\alpha$ elements can be observed \citep{zasowski2019}.

In the top panels of Figure \ref{fig:nonalfa_apo_gse}, we show the abundances for Al and Na. In the APOGEE field samples, we can observe the difference in the distribution of Al and Na in the stellar populations of the Galaxy. The chemical profiles of [Al/Fe] (top left) shows a clear distinction between the APOGEE outer disk sample (gray) and the isodensity contours associated with GSE (yellow), while the [Na/Fe] chemical profiles show a higher dispersion with substantial overlap between both field populations.

The distinction mentioned between the APOGEE field samples (outer disk vs. GSE) can be observed for our TriAnd candidates in the [Al/Fe] ratio. Although an apparent offset between our derived abundances and  the APOGEE ones is present, it is possible to observe a clear difference in the [Al/Fe] vs. [Fe/H] relation between the stars with \textit{in situ} characteristics and the high-eccentricity (likely accreted) ones. This offset has already been mentioned in the literature (e.g., \citealt{griffith2019}) when comparing chemical abundance data for the odd-$Z$ elements in the APOGEE database with the GALAH database.

The [Na/Fe] presented in Figure \ref{fig:nonalfa_apo_gse} for the TriAnd stars presents a dispersion (${\sim}0.25$ dex) very similar to the chemical profiles observed by \cite{bergemann2018}, reinforcing the \textit{in situ} characteristics of the TriAnd population. On the other hand, the suggested accreted stars from our sample occupy the same region as the contours associated with GSE and present a higher overall dispersion in [Na/Fe] and metallicity than the TriAnd stars.

As already mentioned, SNe Ia formed by low-mass stars are the main sources of interstellar medium enrichment for the iron-peak elements. The MW \textit{in situ} populations are expected to present chemical characteristics linked to an extended star-formation history, where the interstellar medium enrichment lasts for long time periods ($\sim$1 Gyr; \citealt{matteucci1989}). Galaxies with extended star formation histories have higher [Ni/Fe] and [Cr/Fe] abundances \citep{kirby2019, larsen2022}.

The chemical profiles from iron-peak elements of our sample can be observed in the bottom panels of Figure \ref{fig:nonalfa_apo_gse}. The TriAnd stars follow the same profile as the outer disk population, showing a small dispersion, as can be seen in the APOGEE data \citep{hayes2018}.

The [Cr/Fe] (bottom left panel of Figure \ref{fig:nonalfa_apo_gse}) shows only a minor overlap with stars from our TriAnd sample with the GSE contours, at the edge of the distribution. On the other hand, the accreted selection essentially overlaps the GSE position.

In the bottom right panel of Figure \ref{fig:nonalfa_apo_gse}, the [Ni/Fe] presents a clear distinction between the APOGEE outer disk sample and the GSE contours, with no overlap between these samples. The TriAnd stars and stars with a suggested accreted origin in our sample also present very distinct characteristics. This behavior reinforces the scenario where the TriAnd region is formed by a majority of \textit{in situ} stars and a small contribution of \textit{ex situ} interlopers.

\subsubsection{Neutron-capture elements}

For elements heavier than Fe (atomic number $Z > 26$), nucleosynthesis typically takes place via neutron-capture processes. We can distinguish two major processes: the ``slow'' ($s$-process) and ``rapid'' ($r$-process) neutron-capture processes (see \citealt{sneden2008} for a review).

The traditional main $s$-process produces elements from Sr ($Z = 38$) to Pb ($Z = 82$) majorly in the H-rich intershell of low-mass AGB stars \citep{busso1999, herwig2005, karakas2014}. For the occurrence of $r$-process nucleosynthesis, highly energetic events are expected, such as binary compact mergers \citep{lattimer1974, rosswog1999, wanajo2009, drout2017, thielemann2017, thielemann2020, cowan2021} and/or magnetorotational supernovae \citep{winteler2012,mosta2014,reichert2021}, synthesizing heavy elements with two major peaks around Xe ($Z = 54$) and Pt ($Z = 78$).

\begin{figure*}
    \centering
    \includegraphics[width=1.95\columnwidth]{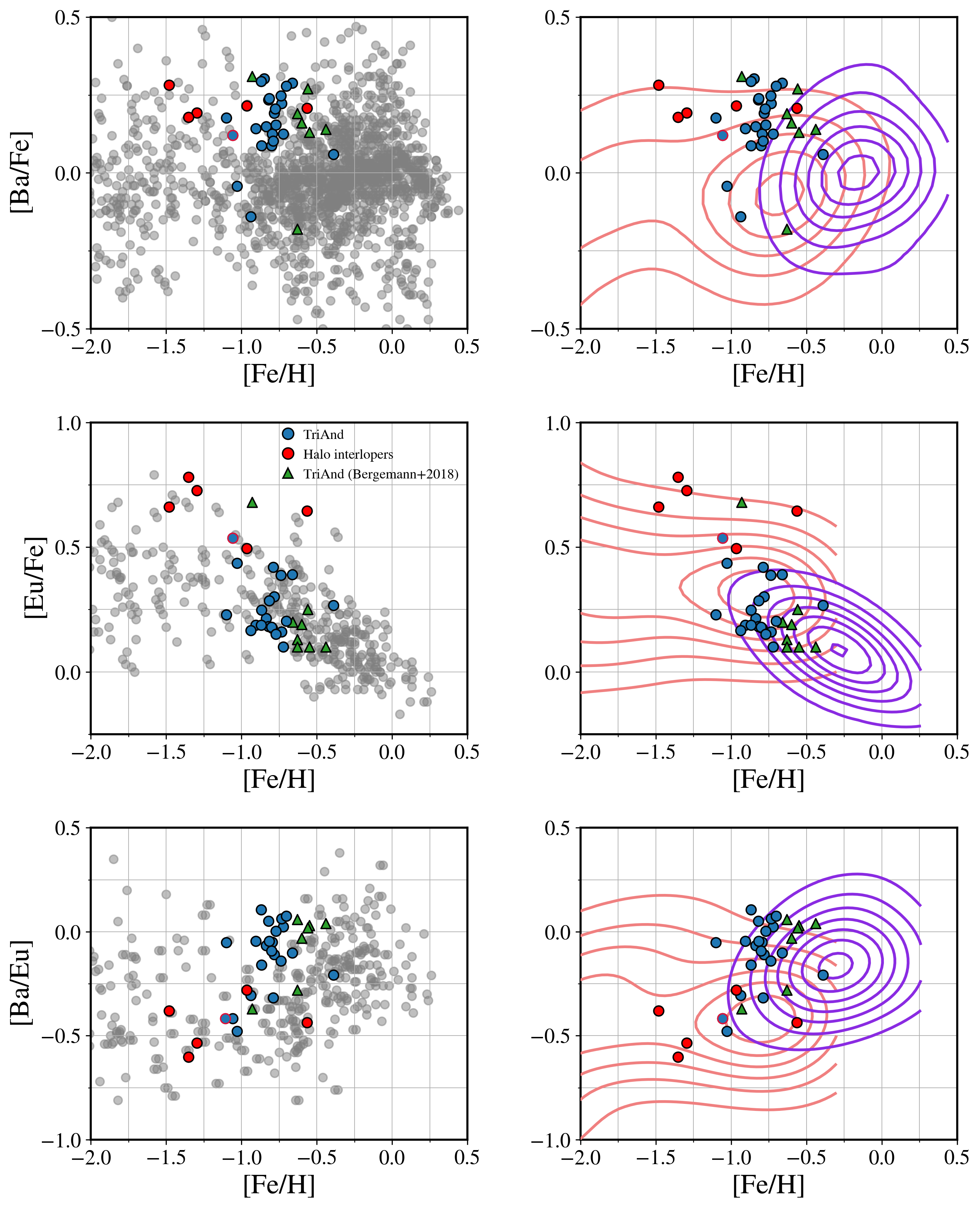}
    \caption{The [X/Fe] and [Ba/Eu] ratio vs. [Fe/H] for the neutron-capture elements (Ba and Eu), with each chemical element depicted in its respective panel. The TriAnd candidate stars employed in this work are represented by blue circles, the eccentricity selection is represented by red circles, the \cite{bergemann2018} sample is represented by green triangles, field stars from the SAGA database \citep{saga2008, saga2017} are in gray, and isodensity contours associated with the thin and thick disk in are purple and coral, respectively.}
    \label{fig:baeu_saga}
\end{figure*}

In Figure \ref{fig:baeu_saga}, the abundances of Ba and Eu are presented, elements representative of the $s$- and $r$-processes, respectively, within the metallicity range considered ($-1.5 < \rm[Fe/H] < -0.5$). For comparison with our TriAnd sample, we obtained data from the Stellar Abundances for Galactic Archaeology (SAGA) database \citep{saga2008,saga2017}, shown as gray dots, since the APOGEE DR17 database does not contain Ba and Eu abundances. We could not apply the same outer disk selection as presented in Figures \ref{fig:alfa_apo_gse} and \ref{fig:nonalfa_apo_gse} as the stars from the SAGA database do not reach large $R_{\rm GC}$. Therefore, we display all data for Ba and Eu for MW stars contained in this database, excluding upper limits. We also display isodensity contours representing the thin ($\rm[Mg/Fe] < 0.2$) and thick ($\rm[Mg/Fe] \geq 0.2$) disks \citep{li2018,mackereth2019,beraldo2021,myeong2022,queiroz2023} in purple and coral, respectively.

\cite{ratcliffe2022} analyzed the chemical trends of the MW disk in different Galactocentric distances. Their results for the neutron-capture elements presented the largest variation from all of the analyzed elements, showing that the evolution and enrichment sites of these heavy elements do not follow a simple trend along the disk. Indeed, in the top panels of Figure \ref{fig:baeu_saga}, it is shown that both the [Ba/Fe] from this work and the disk stars from the SAGA database present a high dispersion, making it difficult to distinguish between different stellar populations. 

Even amidst a high dispersion, the majority of our calculated abundances can be found in the range 0.1 < [Ba/Fe] < 0.3. These high values of [Ba/Fe], when compared to the local disk stars from SAGA database in the same metallicity range, agree with the works from \cite{bergemann2018} and \cite{sales2019} for their sample of TriAnd candidates. This scenario can probably be explained by the positive gradient from the $[s/Fe]$ ratio present in the Galactic disk. The TriAnd stellar population is located in a higher $R_{\rm GC}$, presenting a higher $[s/Fe]$ ratio than stars in the local disk (see, e.g. \citealt{sales2022}).

Contrary to the [Ba/Fe], the [Eu/Fe] ratio, presented in the middle panels of Figure \ref{fig:baeu_saga}, decreases with an increasing metallicity, as expected for $r$-process elements that are linked to high-mass stars. From our calculated [Eu/Fe] abundances, we can observe that the majority of our sample is in good agreement with the disk stars from the SAGA database and \cite{bergemann2018} sample, favoring an \textit{in situ} origin.

Conversely, the high-eccentricity stars present chemical patterns similar to accreted populations, usually presenting an $r$-process enrichment ($\rm[Eu/Fe] \geq 0.5$). Interestingly, we also identified that the three most metal-poor stars in our sample ($-1.50 < \rm[Fe/H] < -1.25$) are possible $r$-II stars \citep{beers2005}, a class of highly $r$-process enhanced stars that are thought to have originated in rare, neutron-rich sites, presenting low [Ba/Eu] (${<}0$) and high [Eu/Fe] (${\gtrsim} +0.7$) values \citep{holmbeck2020}.

In the bottom panels of Figure \ref{fig:baeu_saga} we can observe the [Ba/Eu] ratio; these abundances reflect the relative number of high- to low-mass stars enriching the interstellar medium where these stars were formed. Even with a high dispersion, we can observe that the majority of our true TriAnd stars present an overall higher [Ba/Eu] ratio, meaning a predominant enrichment from the $s$-process over the $r$-process. Alternatively, the stars with a possible \textit{ex situ} origin show lower [Ba/Eu], whereas our most metal-poor stars with accreted characteristics present [Ba/Eu] ${\sim} -0.5$ expected for an \textit{ex situ} population \citep{aguado2021,aguado2021b,limberg2021b,matsuno2021,ji2022,naidu2022}.

\section{Summary}
\label{sec:conclusion}
Divergent suggestions about the nature of TriAnd haven been raised since its discovery. In this work, taking advantage of the largest homogeneous sample of TriAnd candidate stars analyzed with high-resolution spectra, we performed a chemodynamical investigation of an expanded sample of 31 TriAnd candidate stars in order to better understand the origin of its stellar population.

From the orbital parameter analysis of our TriAnd candidates, we observed that the majority of our sample falls within the range of the orbital parameters typical of stars from the Galactic disk. Moreover, through an eccentricity cut ($e > 0.4$), we identified that the subsample with higher eccentricity presents similar properties to MW accreted populations, such as GSE, indicating an accreted origin for these stars.

The abundance analysis identified that the majority of our TriAnd candidates present chemical patterns similar to the outer thin-disk population. Reassuringly, out of our TriAnd members, those chemically more akin to an \textit{in situ} nature are specifically those on low-eccentricity orbits ($e < 0.4$). On the contrary, the high-eccentricity subsample exhibits an abundance profile similar to accreted MW populations, such as GSE.

Our chemodynamical study indicates an \textit{in situ} origin for TriAnd, as the majority of the sample analyzed in this work presents properties, both dynamical and chemical, similar to the outer thin-disk population. We also attributed the suggested ``knee” pattern in the relation between the [$\alpha$/Fe] ratio and [Fe/H] as a contamination of \textit{ex situ} stars at the same distance and location of TriAnd in past literature samples.  Finally, our analysis strongly suggests that the contradictory interpretations found in the literature about the origin of such an overdensity, including extragalactic (e.g., \citealt{chou2011, deason2014}), \textit{in situ} (e.g., \citealt{bergemann2018, hayes2018}), and ``unique" (e.g., \citealt{sales2019, sales2020}), were mainly due to the smaller number of stars available for analysis in previous works.

\acknowledgments
Y.A. would like to acknowledge financial support from CAPES (proc. 88887.604784/2021-00). H.D.P. thanks FAPESP (proc. 2018/21250-9 and 2022/04079-0). S.R. would like to acknowledge partial financial support from FAPESP (procs. 2015/50374-0 and 2014/18100-4), CAPES, and CNPq. G.L. acknowledges FAPESP (procs. 2021/10429-0 and 2022/07301-5). A.P.-V. acknowledges DGAPA-PAPIIT grant IA103122. R.M.S. acknowledges CNPq (proc. 306667/2020-7). The work of V.M.P. is supported by NOIRLab, which is managed by the Association of Universities for Research in Astronomy (AURA) under a cooperative agreement with the National Science Foundation. J.V.S.-S. acknowledges support from FAPERJ under grant 203.899/2022. F.A. acknowledges financial support from MICINN (Spain) through the Juan de la Cierva-Incorporación program under contract IJC2019-04862-I. Y.A and H.D.P. thank Katia Cunha for its support with the proposal and discussions in the early phase of the project. Y.A. and S.R. thank Chris Sneden for the helpful insights about using \texttt{MOOG} and how to make a better spectral synthesis; Henrique Reggiani and Jhon Galarza for discussions about the spectral analysis; the Gemini help desk team, especially Eder Martioli, for help with GRACES data and reduction, and Ulisse Munari for providing the synthetic spectra. This work was enabled by observations made from the Gemini North telescope, located within the Maunakea Science Reserve and adjacent to the summit of Maunakea. We are grateful for the privilege of observing the universe from a place that is unique in both its astronomical quality and its cultural significance. Based on observations obtained through the Gemini Remote Access to CFHT ESPaDOnS Spectrograph (GRACES). ESPaDOnS is located at the Canada-France-Hawaii Telescope (CFHT), which is operated by the National Research Council of Canada, the Institut National des Sciences de l’Univers of the Centre National de la Recherche Scientifique of France, and the University of Hawai’i. ESPaDOnS is a collaborative project funded by France (CNRS, MENESR, OMP, LATT), Canada (NSERC), CFHT, and ESA. ESPaDOnS was remotely controlled from the international Gemini Observatory, a program of NSF’s NOIRLab, which is managed by the Association of Universities for Research in Astronomy (AURA) under a cooperative agreement with the National Science Foundation on behalf of the Gemini partnership: the National Science Foundation (United States), the National Research Council (Canada), Agencia Nacional de Investigación y Desarrollo (Chile), Ministerio de Ciencia, Tecnología e Innovación (Argentina), Ministério da Ciência, Tecnologia, Inovações e Comunicações (Brazil), and Korea Astronomy and Space Science Institute (Republic of Korea).
IRAF was distributed by the National Optical Astronomy Observatory, which was managed by the Association of Universities for Research in Astronomy (AURA) under a cooperative agreement with the National Science Foundation.

\facilities{Gemini:North (GRACES)}

\software{
          {\tt OPERA} \citep{martioli2012},
          {\tt IRAF} \citep{iraf},
          {\tt MOOG} \citep{moog},
          {\tt qoyllur-quipu} \citep{q2},
          {\tt matplotlib} \citep{matplotlib},
          {\tt numpy} \citep{numpy},
          {\tt pandas} \citep{pandas},
          {\tt scipy} \citep{scipy},
          {\tt StarHorse} \citep{starhorse2}.
          }

\clearpage

\bibliographystyle{aasjournal}

\bibliography{bibliography.bib}

\end{document}